\documentstyle[12pt]{article}
\begin{document}
\thispagestyle{empty}
\def\f{phonoriton\ }
\def\fs{phonoritons\ }
\begin{center}

\null
\vskip-1truecm
\rightline{IC/95/226}
\vskip1truecm International Atomic Energy Agency\\ and\\ United Nations
Educational Scientific
and Cultural Organization\\
\medskip INTERNATIONAL CENTRE FOR THEORETICAL PHYSICS\\
\vskip1.5truecm
{\bf  DENSITY-DEPENDENT PHONORITON STATES \\
IN HIGHLY EXCITED SEMICONDUCTORS}\\
\vspace{1cm}
 Nguyen Hong Quang\footnote{\normalsize On leave of absence from:
Institute of Theoretical Physics, P.O. Box 429, Hanoi 10000, Vietnam.}\\
International Centre for Theoretical Physics, Trieste, Italy.\\[0.5cm]
 Nguyen Minh Khue\\ Institute of Theoretical Physics,
P.O.Box 429 Bo Ho, Hanoi 10000, Vietnam
\end{center}
\vspace{0.5cm}
\centerline{ABSTRACT}
\bigskip
\baselineskip 0.7cm

The dynamical aspects of the phonoriton state in highly-photoexcited
semiconductors is studied theoretically. The effect of the
exciton-exciton interaction and nonbosonic character of high-density
excitons are taken into account.  Using Green's function method and
within the Random Phase Approximation it is shown that the phonoriton
dispersion and damping are very sensitive to the exciton density,
characterizing the excitation degree of semiconductors.\\

\vspace{1.5cm}
\begin{center}
MIRAMARE -- TRIESTE\\
\medskip
August 1995
\end{center}

\newpage
\noindent
{\bf 1. INTRODUCTION}

It is expected that under an intense electromagnetic radiation near
the exciton resonance, the interaction between excitons and phonons
can lead to the formation of a new kind of elementary excitation in
semiconductors called \f [1-13].  In treating the problem all the
authors in [1-13] have considered excitons as ideal bosons and in some
cases, have neglected the exciton-exciton interaction, although as
mentioned therein, dealing with phonoriton problem, in fact we have
matter with high-excitation phenomena, where these assumptions are no
more valid. It is well known that in highly-excited semiconductors
the so-called giant polariton states are drastically different from
that of the usual Hopfield polariton. In such a condition when the
number of excitons is very large, the excitons cannot be considered
as ideal bosons, and the interaction between them becomes
considerable and cannot be neglected. The nonbosonic character of
excitons and the exciton-exciton interaction can lead to many
interesting effects in semiconductors [14-16]. As a mixture of
excitons, photons and phonons, the \f spectrum and damping must
depend on the degree of excitation. However, up to now there has been
no papers concerning this aspect of the phonoriton problem.

The purpose of our work is to develop an approach to study the
dynamical aspects of the \f reconstruction  in highly-excited
semiconductors. For simplicity, we shall use the simple two-level
model recently developed by Wang and Birman [11] to consider both the
above-mentioned effects on the \f state. As will be seen the \f
spectrum and damping are very sensitive to the exciton density
characterizing the excitation degree of semiconductors.

  We use the unit system with $\hbar = c = 1$.\\[0.5cm]
{\bf 2. THE EFFECTIVE HAMILTONIAN}

Let us consider a semiconductor with direct band gap illuminated by
high-intensity pump field resonant to the lowest 1S exciton state. In
such condition the occupation number for the exciton in mode $k$ with
frequency equal to that of the incident field, will be very  large
(mode $k$ is macroscopically occupied). The creation and annihilation
operators of the exciton $a^+_k , a_k $ satisfy the following exact
commutation relations [16]
\begin{eqnarray}
\left[ a_k , a_{k'} \right] &=&\left[ a^+_k , a^+_{k'} \right] = 0 \
\ ,\nonumber \\[0.3cm]
\left[ a_k , a^+_{k'} \right] &=& \delta_{k, k'} - \frac{1}{V} \sum_p
\left\{ \varphi^*(p-\beta k) \varphi(p-\beta k') e^+_{k'-p} e_{k-p} \
\ \ + \right. \nonumber \\[0.3cm]
&+& \left. \varphi^*(p+\alpha k) \varphi(p+\alpha k')
h^+_{k'+p} h_{k+p}  \right\} ,
\end{eqnarray}
where $e^+_p (e_p ) , h^+_p ( h_p )$ are the creation (annihilation)
operators for the electron and hole with momentum $p$ ,
respectively. $\varphi (p) $ is the Fourier transform of the exciton
envelope function for electron-hole relative motion.  $\alpha = m_e /
(m_e +m_h) = 1 - \beta$, $m_e$ and $m_h$ are the effective masses of
the electron and hole, respectively. $V$ is the volume of the
crystal. From (1) it is clear that the exciton can be considered as
ideal boson only at low excitation, when the second and third terms
on the right-hand side of (1) are neglected.

In addition to the Hamiltonian (1) in Ref.[11] we must also include
the term describing the nonlinear exciton-exciton interaction. So our
initial Hamiltonian has the following form:
\begin{eqnarray}
H & = & H^{ex-\gamma} + H^{ex-ph} \ \ , \\[0.3cm]
H^{ex-\gamma} & = & \omega^{ex}_k a^+_k a_k + \omega^{\gamma}_k b^+_k
b_k + g_k ( a^+_k b_k + b^+_k a_k ) + f_k a^+_k a^+_k a_k a_k \ \ ,
\\[0.3cm]
H^{ex-ph} & = & \sum_p \Omega_p c^+_p c_p + \sum_{p,q} i M(p-q) a^+_p
a_q ( c_{p-q} - c^+_{q-p} ) \ \ ,
\end{eqnarray}
where $a^+_k (a_k) , b^+_k (b_k) , c^+_k (c_k)$ are creation
(annihilation) operators for nonbosonic excitons, photons and
phonons, respectively; $\omega^{ex}_k$ is the energy of the exciton
with momentum $k; \omega^{ex}_k = E_t + k^2/2m_{ex} ( E_t$  is the
exciton energy at $k=0$, $m_{ex}$ is the effective mass of exciton),
$\omega^\gamma_k$ is the energy of the photon , $\Omega_p$ is the
energy of the optical phonon, $g_k$ is coefficient of exciton-photon
interaction, $M(p-q)$ is the matrix element of exciton-phonon
interaction, $f_k$ denotes the constant of exciton-exciton
interaction. Note that unlike [11] $a^+_k$ and $a_k$ stand here for
nonbosonic excitons and their commutators satisfy (1).  Due to this
nonbosonity of the excitons $H^{ex-\gamma}$ in (2) can not be
diagonalized  by the Bogoliubov transformation like it was done in
[11-13]. To overcome this difficulty we introduce the following
Green's functions
\begin{eqnarray}
G_1 (k , t) & = & \theta(t) < [a_k(t) , a^+_k(o)] > \ \ \equiv \ \ <<
a_k(t) , a^+_k(o) >> \ , \nonumber \\[0.3cm]
G_2 (k,t) & \equiv & << b_k(t) , a^+_k (o) >> \ ,
\end{eqnarray}
where the symbol $< \ldots > $ denotes the thermodynamical average.

The equations of motion for the Green's functions (5) can be derived
in standard way. In order to make the obtained system of equation
closed we use RPA-type factorization [17] which seems to be
reasonable for high density systems where the fluctuation of the
exciton densities can be neglected. As a result we obtain the closed
system of equation for Green's functions in energy representation
\begin{eqnarray}
\left[\omega^{pol}_k - \omega^{ex}_k (1 - \rho_k ) - 2f_k n_{ex} (1 -
\rho_k ) \right] G_1(k) - g_k \ ( 1 - \rho_k ) G_2(k) & = &
\frac{1}{2\pi}(1 - \rho_k),\nonumber \\[0.3cm]
g_k G_1(k) - (\omega^{pol}_k - \omega^\gamma_k) G_2(k) & = & 0 \ \ ,
\end{eqnarray}
where $n_{ex}$ is the exciton density in the semiconductors, and
\begin{equation}
\rho_k = \sum_p \left\{|\varphi (p-\beta k)|^2 n_e(k-p) +
|\varphi(-p+\alpha k)|^2 n_h (k-p) \right\} \ \ ,
\end{equation}
$n_e(k), n_h(k)$ are the electron and hole densities, respectively.

Solving (6) we obtain the dispersion of the polariton {\it i.e.}
dressed exciton as a pole of the Green's function in the following
form
\begin{eqnarray}
\omega^{pol}_{i,k} &=& \frac{1}{2} \left[(\omega_k^{ex} + 2n_{ex}
f_k)(1 - \rho_k) + \omega^\gamma_k \right]\quad - \nonumber \\[0.3cm]
&-& \frac{(-1)^i}{2}\left\{\left[ (\omega^{ex}_k + 2n_{ex} f_k )(1 -
\rho_k) -  \omega^\gamma_k \right]^2 + 4(1 - \rho_k) g^2_k
\right\}^{\frac{1}{2}} \ \ ,
\end{eqnarray}
$i = 1, 2 . $

In order to study the phonoriton state it is convenient for us to
work with bosons rather than nonbosonic particles. This idea can be
realized if we construct the following effective Hamiltonian for the
exciton-photon part in term of boson-excitons $\tilde a_k$ so that
this Hamiltonian gives us the same polariton dispersion (8)
\begin{equation}
H^{ex-\gamma}_{eff} = \tilde{\omega}^{ex}_k \tilde{a}^+_k
\tilde{a}_k+ \omega^\gamma_k b^+_k b_k + \tilde g _k (\tilde{a}^+_k
b_k + b^+_k \tilde{a}_k ) .
\end{equation}

Indeed, if we chose $\tilde\omega^{ex}_k$ and $\tilde g_k$ , as
follows
\begin{eqnarray}
\tilde\omega^{ex}_k & = & ( \omega^{ex}_k + f_k 2 n_{ex} ) ( 1 -
\rho_k ), \nonumber \\[0.3cm]
\tilde g_k & = & g_k \sqrt{1 - \rho_k} ,
\end{eqnarray}
we can diagonalize the effective Hamiltonian (9) by a canonical
Bogoliubov transformation to get it in the form
\begin{equation}
H^{ex-\gamma}_{eff} = \sum_{k,i} \omega^{pol}_{i,k} B^+_{ik}B_{ik} ,
\end{equation}
where $\omega^{pol}_{i,k}$ is exactly defined by (8). $B^+_{ik} $ is
the creation operator for the polariton in i-branch and with momentum
$k$, and expressed in terms of photon ($b_k$) and boson-exciton
($\tilde{a}_k$), as follows
\begin{equation}
B_{ik} = u_i(k) b_k + v_i(k) \tilde a_k ,
\end{equation}
where
\begin{eqnarray}
u_i(k) &=& \frac{\tilde g_k}{\omega^{pol}_{ik} - \omega^\gamma_k}
v_i(k), \nonumber \\[0.3cm]
v_i(k) &=& \frac{|\omega^{pol}_{ik} - \omega^\gamma _k |}{\{
(\omega^{pol}_{ik} - \omega^\gamma_k)^2 + \tilde g_k^2 \}^{1/2}} .
\end{eqnarray}

Substituting (11) and (12) into (2), and following the same arguments
as in [11] we obtain the final effective Hamiltonian for the
anti-Stokes scattering of the incident macroscopically filled
polariton mode $k_o$ of lower polariton branch.
\begin{eqnarray}
H_{AS} &=& \sum_p (\omega^{pol}_p + i\gamma^{pol}_{p}) B^+_pB_p +
\sum_p (\Omega_p + i\gamma^{ph}_p) c^+_p c_p \ +\nonumber \\[0.3cm]
&+& \sum_{p \ne k_o}i \tilde M(p-k_o) ( B^+_pB_{k_o}c_{p-k_o} -
B^+_{k_o} B_p c^+_{p - k_o} ) ,
\end{eqnarray}
where
\begin{eqnarray}
\omega^{pol}_p &\equiv &\omega^{pol}_{2, p} \ \ , \nonumber \\[0.3cm]
\tilde M (p-k_o) &\equiv & M(p-k_o) v^*_2(p) v_2(k_o) .
\end{eqnarray}

Note that the Hamiltonian (14) formally looks like Hamiltonian (12)
in [11], however, there is an essential difference between them.
While the dispersion of the polariton $\omega ^{pol}_p$ in [11] is
fixed in respect to external excitation, our $\omega ^{pol}_p$ is
function of exciton density characterizing the degree of excitation.
This is a dynamical consequence of the nonlinear exciton-exciton
interaction and the nonbosonity of high density excitons.  Moreover,
we have also included lifetime effects introducing $\gamma^{pol}_p$
and $\gamma^{ph}_p$ - the damping of polariton and phonon,
respectively. \\[0.5cm]
{\bf 3. PHONORITON SPECTRUM AND DAMPING}

In this section we shall use the effective Hamiltonian derived in the
previous section to study phonoriton states in highly excited
semiconductors.

Starting now from (14) one can follow [11] replacing $B_{k_o},
B^+_{k_o}$ by their c-number expectation values $<B_{k_o}>$ and $<
B^+_{k_o}>$ to get phonoriton dispersion. However, as shown in our
previous work [13] such a semiclassical approximation may omit many
interesting properties of phonoritons. So in this paper we shall use
the Green's function method [17], which is more reasonable and
consistent for us in the framework of the RPA. Thus, let us consider
the following Green's functions
\begin{eqnarray}
F_1(p,t) &=& << B_p(t) , B^+_p(0) >> \ \ , \nonumber \\[0.3cm]
F_2(p,t) &=& << B_{k_o}(t)c_{p-k_o}(t) , B^+_p(0) >> .
\end{eqnarray}

The equations of motion for the Green's functions (16) can be derived
in standard way. It is easy to check that in the framework of the RPA
the Fourier images of the Green's functions (16) are the solutions of
the following closed system of equations
\begin{eqnarray}
(\omega - \omega^{pol}_p -i \gamma^{pol}_p) F_1(p,\omega) - i \tilde
M (p-k_o) F_2 (p,\omega)
&=& \frac{1}{2 \pi}, \nonumber \\[0.3cm]
i \tilde M (p - k_o) n_{ex} V F_1 (p,\omega) + (\omega -
\omega^{pol}_{k_o} - i \gamma^{pol}_{k_o} - \Omega_{p-k_o} -i
\gamma^{ph}_{p-k_o})  F_2 (p,\omega) & = & 0 ,
\end{eqnarray}
where we have approximately replaced the occupation number $N_{k_o} =
< B^+_{k_o} B_{k_o} >$ of the polariton at mode $k_o$
by $ N_{ex} = V.n_{ex} $ due to the exciton-like nature of the
polariton mode $k_o$ near the exciton resonance.

{}From the solutions of the system of equations (17) we extract the
poles of the functions $F_1(p,\omega)$ and $ F_2(p,\omega)$, the real
and imaginary parts of which, as usual, give us the spectrum and
damping of the new quasiparticle - the phonoriton.

For the spectrum of phonoriton we obtain the following final
expression
\begin{eqnarray}
W_{i}(p) &=& \frac{1}{2}(\omega^{pol}_p +\omega^{pol}_{k_o} +
\Omega_{p-k_o}) + \nonumber \\[0.3cm]
&+& (-1)^{i+1} \frac{1}{2} \left\{ \frac{1}{2} \left[ D(p)^2 +
4\tilde M^2(p-k_o)Vn_{ex} -\gamma^2(p) \right] \right.+ \nonumber
\\[0.3cm]
&+& \left. \frac{1}{2}\sqrt{ \left[ D(p)^2 + 4\tilde
M^2(p-k_o)Vn_{ex} -\gamma^2(p) \right]^2 + 4 D(p)^2 \gamma^2(p)}
\right\}^{\frac{1}{2}} ,
\end{eqnarray}
and for the damping of phonoriton we have
\begin{eqnarray}
\Gamma_i(p) &=& \frac{1}{2} \Gamma(p) + (-1)^{i+1} \frac{1}{2} D(p)
\gamma(p)\left\{ \frac{1}{2} \left[ D(p)^2 + 4\tilde
M^2(p-k_o)Vn_{ex} -\gamma^2(p) \right] + \right. \nonumber \\[0.3cm]
&+& \left. \frac{1}{2}\sqrt{ \left[ D(p)^2 + 4\tilde
M^2(p-k_o)Vn_{ex} -\gamma^2(p) \right]^2 + 4 D(p)^2 \gamma^2(p)}
\right\}^{- \frac{1}{2}} ,
\end{eqnarray}
where
\begin{eqnarray}
D(p) &=& \omega^{pol}_p - \omega^{pol}_{k_o} - \Omega_{p-k_o} ,
\\[0.3cm]
\gamma(p) &=& \gamma^{pol}_p - \gamma^{ph}_{p-k_o} ,\\[0.3cm]
\Gamma(p) &=& \gamma^{pol}_p + \gamma^{ph}_{p-k_o} ,
\end{eqnarray}
\begin{eqnarray}
 \omega^{pol}_p  &=& \frac{1}{2}\left[ (\omega^{ex}_k + 2 f_k n_{ex}
) ( 1 - \rho_k ) + \omega^\gamma _k \right] \ \ - \nonumber \\[0.3cm]
&-& \frac{1}{2}\left\{ \left[ (\omega^{ex}_k + 2 f_k n_{ex} ) ( 1 -
\rho_k ) - \omega^\gamma _k \right]^2 + 4 g^2_k (1 - \rho_k)
\right\}^{\frac{1}{2}} \ .
\end{eqnarray}

\noindent $\rho_k \approx \lambda \pi . a^3_B. n_{ex} $ [16], where
$\lambda$ is some constant parameter. The formulas (18) and (19) are
the final expressions for the dispersion and damping of the
phonoriton. As we see, these analytical expressions depend explicitly
on the exciton density that characterises the level of excitation.
{}From (18) we have a new gap of phonoriton
\begin{eqnarray}
\Delta (p) &=& \left\{ \frac{1}{2} \left[ D(p)^2 + 4\tilde
M^2(p-k_o)Vn_{ex} -\gamma^2(p) \right] \right. + \nonumber \\[0.3cm]
&+& \left. \frac{1}{2}\sqrt{ \left[ D(p)^2 + 4\tilde
M^2(p-k_o)Vn_{ex} -\gamma^2(p) \right]^2 + 4 D(p)^2 \gamma^2(p)}
\right\}^{\frac{1}{2}}
\end{eqnarray}

It is clear from (18)-(19) that the phonoriton dispersion and damping
depend on exciton density not only via  $4\tilde M^2(p-k_o)Vn_{ex}$
as in [10-12], but also through the polariton dispersion (23). If we
put $f_k = 0 $ and $\rho_k = 0 $ we reobtain the results of [11].

To see the sensibility of phonoriton states to the exciton density,
in Fig.1, 2, and 3  we have plotted the phonoriton dispersion, its
gap and damping in CdS for some exciton densities, respectively. The
damping of polaritons can be calculated via the damping of excitons
and photon (see for example [18]), and $f = 13 \pi E_B a^3_B /3V $
[19], where $a_B$ is the effective Bohr radius of the exciton and
$E_B$ is its binding energy.  The data used for numerical
calculation are adopted from [11]: $E_t = 2553 $mev, $ g_o = 50 $mev,
$a_B = 2.8\times 10^{-7} $cm, $E_B = 27.8 $mev, $k_o = 3.98\cdot
10^5$ cm$^{-1}, m_{ex} = 0.98 m_o, \gamma^{ex} = 0.1$meV,
$\gamma^{ph} = 0.05$meV.  Note that the higher exciton density the
lower two phonoriton branches  go down.  With increasing exciton
density not only the phonoriton gap becomes larger as in [11], but
also the phonoriton resonance region shifts to higher values of
momentum. This situation may affect experimental detection of
phonoriton and must be taken into account.\\[0.5cm]
{\bf 4. CONCLUSION}

In this paper we study for the first time the dynamical aspects of
the phonoriton states in highly excited semiconductors. Unlike the
previous papers on phonoritons considering the polariton as fixed in
respect to external excitation before it comes to interaction with
phonon, we have shown that the polariton itself depends on
excitation. The intrinsic nonlinear interactions to high-density
systems such as nonlinear exciton-exciton interaction and nonbosonic
nature of excitons are taken into account. As consequence, these
nonlinear interactions lead to dynamical reconstruction of phonoriton
states. At fixed $k_o$-mode of photon excitation, the higher exciton
density (the higher excitation degree) the deeper the phonoriton
resonance region goes down, and the larger values of momentum $k$ the
latter shifts to. Note that if we neglect these interactions, {\it
i.e.} $f_k \rightarrow 0$ and $\rho_k \rightarrow 0$ all our results
return to the previous ones of [11]. The sensibility of the
phonoriton spectrum and damping upon the degree of external
excitation usually forgotten up to now should be taken into account,
especially in experimental detection of phonoritons. And we hope that
our specific predictions will stimulate further experiments.\\[1.5cm]
{\bf Acknowledgements}

One of us (N.H.Q) would like to take this opportunity to express his
gratitude to the International Centre for Theoretical Physics,
Trieste, for hospitality.  He would also like to thank Prof. B.F.
Zhu for a reading of the manuscript  and discussions, and Prof. Yu Lu
for encouragements.
\newpage
{\bf REFERENCES}

\begin{enumerate}
\itemsep -.1cm
\item A.L. Ivanov and L.V. Keldysh,   Dokl. Acad. Nauk.USSR,  {\bf
264}, 1363 (1982)
\item A.L. Ivanov and L.V. Keldysh,  Zh. Eksp. Teor. Fiz. {\bf 84},
404 (1983)
\item A.L. Ivanov, Zh. Eksp. Teor. Fiz. {\bf 90}, 158 (1986)
\item L.V. Keldysh and S.G. Tikhodeev,   Zh. Eksp. Teor.  Fiz. {\bf
90}, 1852 (1986)
\item L.V. Keldysh and S.G. Tikhodeev, Zh. Eksp. Teor. Fiz. {\bf 91},
78 (1986)
\item G.S. Vyskoviskii, G.P. Golubev, E.A. Zhukov, A.A. Fomichev, and
M.A. Yaskin, Pis'ma Zh. Eksp. Teor. Fiz. {\bf 42}, 134 (1985)
\item M.S. Brodin, V.N. Kadan and M.G. Matsko, Fiz. Tver. Tela {\bf
30}, 1265 (1988)
\item B.I. Greene, J.F. Mueller, J. Orenstein, D.H. Rapkine, S.
Schmitt-Rink, and M. Thakar, Phys. Rev. Lett. {\bf 61}, 325 (1988)
\item T. Ishihara, J. Phys. Soc. Jpn. {\bf 57}, 2573 (1988)
\item M.I. Schmigliuk and V.N. Pitei, {\it Coherent polaritons in
semiconductors.} Schtiintsa edt., Kishinev, 1989
\item B.S. Wang and J.L. Birman, Solid State Commun. {\bf 75}, 867
(1990); Phys. Rev. B {\bf 42}, 9609 (1990).
\item N.A. Viet, N.Q. Huong, and L.Q. Thong, ICTP's Preprint
IC/92/320 (1992)
\item N.M. Khue, N.Q. Huong and N.H. Quang, J.Phys.: Condens. Matter
{\bf 6}, 3221 (1994)
\item M.L. Steyn-Ross and C.W. Gardiner, Phys. Rev. A {\bf 27}, 310
(1983)
\item N.B. An, Intern. J. Mod. Phys. B{\bf 5}, 1215 (1991)
\item S.A. Moskalenko, {\it Introduction to theory of high density
excitons.}  Schtiintsa edt., Kishinev, 1983
\item A.A. Abrikosov, L.P. Gorkov and I.E. Dzyaloshinski {\it Methods
of quantum field theory in statistical physics}. Prentice-Hall, Inc.,
Englewood Cliffs, Newjersey 1963
\item H.N. Cam, N.V. Hieu and N.A. Viet, Phys. stat. sol.(b) {\bf
126}, 247  (1984)
\item T. Hiroshima, Phys. Rev. B {\bf 40}, 3862 (1989); J. Phys.:
Condens. Matter, {\bf 4}, 3849 (1992)
\end{enumerate}
\newpage
{\bf FIGURE CAPTIONS}

\vspace{2cm}
\noindent
{\bf Fig.1} The dispersion of the anti-Stokes phonoriton at some
exciton densities $n_{ex} = 1\cdot 10^{17}$ cm$^{-3}$ (thick line),
$1.5\cdot 10^{17}$ cm$^{-3}$ (thin line), $2\cdot 10^{17}$ cm$^{-3}$
(dotted line).

\vspace{2cm}
\noindent
{\bf Fig.2} The function $\Delta (p)$ at some exciton densities
$n_{ex} = 1\cdot 10^{17}$ cm$^{-3}$ (thick line), $1.5\cdot 10^{17}$
cm$^{-3}$ (thin line), $2\cdot 10^{17}$ cm$^{-3}$ (dotted line).

\vspace{2cm}
\noindent
{\bf Fig.1} The damping of the anti-Stokes phonoriton at some exciton
densities $n_{ex} = 1\cdot 10^{17}$ cm$^{-3}$ (thick line), $1.5\cdot
10^{17}$ cm$^{-3}$ (thin line), $2\cdot 10^{17}$ cm$^{-3}$ (dotted
line).
\end{document}